\RequirePackage{fix-cm}
\documentclass[a4paper,11pt]{article}

\usepackage{authblk}
\usepackage{amsfonts,amsmath,amssymb,amsthm}
\usepackage{algorithm}
\usepackage{paralist}
\usepackage{booktabs}
\usepackage{url}
\usepackage{hyperref}
\usepackage{graphicx}
\usepackage{microtype}
\usepackage{url}
\usepackage{hyperref}
\usepackage{subcaption}
\usepackage{tikz}
\usepackage{multirow}
\usetikzlibrary{positioning}
\usetikzlibrary{decorations.pathreplacing}
\newcommand{\N}{\mathbb{N}}
\newcommand{\Z}{\mathbb{Z}}
\newcommand{\F}{\mathbb{F}}

\newtheorem{definition}{Definition}
\newtheorem{lemma}{Lemma}
\newtheorem{theorem}{Theorem}

\tikzset{
    mybrace/.style={decorate,decoration={brace,aspect=#1}}
}

\providecommand{\keywords}[1]{\textbf{\textit{Keywords }} #1}

\begin{document}

\title{Building Correlation Immune Functions from Sets of Mutually Orthogonal Cellular Automata}

\author[1]{Luca Mariot}
\author[2]{Luca Manzoni}
	
\affil[1]{{\small Digital Security Group, Radboud University PO Box 9010, Nijmegen, The Netherlands} 
	
	{\small \texttt{luca.mariot@ru.nl}}}

\affil[2]{{\small Dipartimento di Matematica e Geoscienze, Università degli Studi di Trieste, Via Valerio 12/1, Trieste, 34127, Italy}

    {\small \texttt{lmanzoni@units.it}}}

\maketitle

\begin{abstract}
Correlation immune Boolean functions play an important role in the implementation of efficient masking countermeasures for side-channel attacks in cryptography. In this paper, we investigate a method to construct correlation immune functions through families of mutually orthogonal cellular automata (MOCA). First, we show that the orthogonal array (OA) associated to a family of MOCA can be expanded to a binary OA of strength at least 2. To prove this result, we exploit the characterization of MOCA in terms of orthogonal labelings on de Bruijn graphs. Then, we use the resulting binary OA to define the support of a second-order correlation immune function. Next, we perform some computational experiments to construct all such functions up to $n=12$ variables, and observe that their correlation immunity order is actually greater, always at least 3. We conclude by discussing how these results open up interesting perspectives for future research, with respect to the search of new correlation-immune functions and binary orthogonal arrays.
\end{abstract}


\section{Introduction}
\label{sec:intro}
Boolean functions are basic combinatorial objects that map a set of fixed size bitstrings to a single output bit. Notwithstanding their simplicity, such functions find countless applications in many diverse fields of computer science and mathematics~\cite{crama11}. In cryptography, Boolean functions have long been used to design low-level primitives in symmetric ciphers, such as combiners and filters for linear feedback shift registers~\cite{carlet21}. The rationale is that the resilience of these ciphers against different cryptanalytic attacks can be reduced to the cryptographic properties of the underlying Boolean functions. For example, a Boolean function used in the combiner model for Vernam stream ciphers should be at a high Hamming distance from the set of affine functions to resist fast-correlation attacks, or equivalently it should possess a high nonlinearity~\cite{meier88}. At the same time, the output of the function should be statistically independent from any subset of $t$ or fewer variables, to resist correlation attacks of order $t$: in other words, the Boolean functions should be correlation immune of high order $t$~\cite{siegenthaler85}.

The literature related to the cryptographic applications of cellular automata (CA) features a solid body of works investigating the properties of the involved Boolean functions. For example, it has been shown that Wolfram's pseudorandom number generator (PRNG) based on a simple one-dimensional CA is unsuitable for cryptographic purposes, as the underlying local rule 30 is not first-order correlation immune and does not have a high enough nonlinearity~\cite{martin08}. This allows to apply respectively Meier and Staffelbach's correlation attack~\cite{meier91} and Koc and Apohan's inversion attack~\cite{koc97} to efficiently recover the initial configuration of Wolfram's PRNG. Following this research thread, a few subsequent works~\cite{formenti14,leporati14} focused on the search of local rules with a larger diameter and a better trade-off of nonlinearity and correlation immunity. Some of these rules with better properties have then been adopted in the design of CA-based stream ciphers such as CARPENTER and PENTAVIUM~\cite{lakra18,john20}.

The correlation immunity criterion also gained relevance in recent years concerning a different type of attacks, namely \emph{side-channel analysis} (SCA). Instead of focusing on the mathematical design of a cipher as classical cryptanalysis does, SCA targets the \emph{implementation} of a cipher on a device. In particular, the aim of SCA is to exploit leakages on side-channel sources such as electromagnetic emanations, timings, and variations of voltages to infer the secret key used to encrypt a message on the device. One of the options to counteract these attacks is \emph{Boolean masking}, where noise is added to the intermediate values computed by the cipher during the encryption process, changing them from one execution to the other. In this respect, correlation immune functions of high order $t$ and low Hamming weight (i.e. with as few 1s as possible in the output of their truth tables) can be used to implement masking countermeasures, such as leakage squeezing and rotation S-box masking~\cite{carlet12,nassar12}, which have minimal implementation overhead and optimal resistance towards SCA attacks of order $t$.

Contrasting with the large number of works related to the study of the cryptographic properties of CA---be it at the local rule level as mentioned above, or by considering them as S-boxes as in~\cite{mariot18a,mariot19}---there seems to be comparatively a smaller literature dedicated to the exploration of CA as a means to construct side-channel countermeasures. To the best of our knowledge, the works by Karmarkar and Roy Chowdury~\cite{karmakar12,karmakar14,karmakar18} are the only ones addressing the design of leakage squeezing countermeasures through hybrid CA. There, the authors remark that there exist several methods to design linear codes with hybrid CA by leveraging the techniques in~\cite{chaudhuri97}, which can be readily used to implement a leakage squeezing countermeasure. However, they also argue that the linearity of such codes introduces other weaknesses, and thereby set out to study the cryptographic properties of nonlinear hybrid CA for leakage squeezing.

The goal of this paper is to investigate how cellular automata may be used to design correlation immune functions of low weight. To this end, we use a construction of mutually orthogonal CA (MOCA), i.e. a family of CA giving rise to a set of mutually orthogonal Latin squares (MOLS), recently introduced in~\cite{mariot20}. In particular, we exploit the well-known fact that any set of MOLS is equivalent to an orthogonal array (OA) of strength $2$. Then, taking any set of MOCA, we prove that the corresponding binary expansion is a binary OA of strength at least $2$, leveraging on the MOCA characterization as orthogonal labelings on de Bruijn graphs. This result allows us to use the expanded binary OA of a MOCA family as the support of a Boolean function with correlation immunity order at least $2$. We then perform a computational search experiment to generate all families of $k=3$ MOCA defined by local rules of diameter $d=4,5$, and construct the corresponding Boolean functions. Interestingly, all generated functions turn out to have correlation immunity order at least 3, indicating that our theoretical result is not a tight lower bound. Although the Hamming weight reached by our correlation immune functions is far from optimal, we discuss how it could be improved by solving an associated optimization problem. The objective in this case is to remove a subset of rows from the binary expansion of a MOCA family while retaining the correlation immunity order of the resulting function.

The rest of this paper is organized as follows. Section~\ref{sec:bg} collects all necessary background definitions and results related to Boolean functions, cellular automata, Latin squares and orthogonal arrays used in this work. Section~\ref{sec:constr} proves the main theoretical result of the paper, i.e. that the binary expansion of a set of MOCA is the support of a Boolean function with correlation immunity order at least $2$. Section~\ref{sec:search} presents the results of the computational search experiment for families of $k=3$ MOCA, giving rise to Boolean functions of up to $n=12$ variables. Finally, Section~\ref{sec:outro} recaps the main contributions of this paper, and discusses some directions for future research.

\section{Preliminary Definitions}
\label{sec:bg}
In this section, we recall all relevant definitions to describe our results in the remainder of the paper. We start with basic concepts and results of Boolean functions, and how correlation immune functions can be characterized by orthogonal arrays. Then, we give a formal definition of the CA model used in our work, and describe the CA-based construction of mutually orthogonal Latin squares of~\cite{mariot20}.

\subsection{Boolean Functions and Orthogonal Arrays}
\label{subsec:bool}
As a general reference, we follow Carlet's recent book on Boolean functions~\cite{carlet21}.

Let $\F_2 = \{0,1\}$ be the finite field with two elements, with sum and multiplication defined respectively as the XOR (denoted by $\oplus$) and logical AND (denoted by concatenation). For any $n \in \N$, the set $\F_2^n$ of all $n$-bit bitstrings is endowed with the structure of a vector space, with vector sum defined as bitwise XOR, and multiplication by a scalar $a \in \F_2$ being the field multiplication of $a$ with each coordinate of a vector $x \in \F_2^n$. Given two vectors $x, y \in \F_2^n$, their Hamming distance $d_H(x,y)$ is the number of coordinates where $x$ and $y$ disagree, while their scalar product is defined as $x \cdot y = \bigoplus_{i=1}^n x_iy_i$. The support of a vector $x \in \F_2^n$ is the set $supp(x) = \{i: x_i \neq 0\}$, and its Hamming weight $w_H(x)$ is the cardinality of $supp(x)$. Equivalently, $w_H(x)$ corresponds to the Hamming distance $d_H(x,\underbar{0})$ between $x$ and the null vector $\underbar{0} \in \F_2^n$. In practice, the Hamming weight is the number of nonzero coordinates in $x$.

For all $n \in \N$, a \emph{Boolean function} of $n$ variables is a mapping $f: \F_2^n \to \F_2$. The most straightforward way to represent $f$ is by means of its truth table. Suppose that a total order is fixed on the vectors of $\F_2^n$ (e.g., the lexicographic order). Then, the truth table of $f$ is the vector $\Omega_f \in \F_2^{2^n}$ defined as:
\begin{equation}
\label{eq:tt}
\Omega_f = ( f(0,\cdots,0), f(0,\cdots,1), \cdots , f(1,\cdots,1)) \enspace .
\end{equation}
In other words, the truth table is the $2^n$-bit vector that specifies for each input vector $x \in \F_2^n$ in lexicographic order the corresponding output value $f(x)$. Similarly to what we defined above for binary vectors, the support of $f$ is the set $supp(f) = \{x \in \F_2^n: f(x) \neq 0\}$, while its Hamming weight is $w_H(f) = |supp(f)|$. Equivalently, support and weight of $f$ are defined respectively as the set of nonzero coordinates and the size of such set in the truth table $\Omega_f$.

A second common representation of a Boolean function $f: \F_2^n \to \F_2$ is the \emph{Algebraic Normal Form} (ANF). Recalling that $x^2 = x$ for all $x \in \F_2$, the ANF of $f$ is the multivariate polynomial over the quotient ring $\F_2[x]/(x_1^2 \oplus x_1, \cdots, x_n^2 \oplus x_n)$:
\begin{equation}
\label{eq:anf}
P_f(x) = \bigoplus_{u \in F_2^n} a_u x^u = \bigoplus_{u \in F_2^n} a_u (x_1^{u_1}\cdots x_n^{u_n}) \enspace ,
\end{equation}
where $a_u \in \F_2$ for all $u \in \F_2^n$. The \emph{algebraic degree} of $f$ is defined as $deg(f) = \max_{u \in \F_2^n} \{w_H(u): u \neq 0\}$, or equivalently $deg(f)$ is the dimension of the largest nonzero monomial in the ANF of $f$. Boolean functions of degree $1$ are also called \emph{affine}, and in particular an affine function is \emph{linear} if $a_{\underbar{0}} = 0$.

A third method to represent a Boolean function usually adopted in cryptography is the \emph{Walsh transform}. Given $f: \F_2^n \to \F_2$, the Walsh transform of $f$ is the function $W_f: \F_2^n \to \Z$ defined for all $a \in \F_2^n$ as:
\begin{equation}
\label{eq:wt}
W_f(a) = \sum_{x \in \F_2^n} (-1)^{f(x) \oplus a\cdot x}, \enspace .
\end{equation}
Intuitively, $W_f(a)$ measures the correlation between $f$ and the linear function defined by the scalar product $a \cdot x$. The Walsh transform is useful to assess several interesting cryptographic properties of $f$. For example, the nonlinearity property is defined as the Hamming distance of $f$ from the set of all affine functions. Using the Walsh transform, the nonlinearity of $f$ may be computed as:
\begin{equation}
\label{eq:nl}
nl(f) = 2^{n-1} - \frac{1}{2} \max_{a \in \F_2^n}\left\{ |W_f(a)| \right\} \enspace .
\end{equation}
Another cryptographic property which will be of special interest for this paper is correlation immunity. A Boolean function $f: \F_2^n \to \F_2$ is correlation immune of order $1 \le t \le n$ if any subset of at $1 \le k \le t$ input variables is statistically independent from the output of $f$. This property has an equivalent characterization through the Walsh transform (originally due to Xiao and Massey~\cite{xiao88}): $f$ is $t$-th order correlation immune if and only if $W_f(a) = 0$ for all coefficients $a \in \F_2^n$ of Hamming weight $1 \le k \le t$.

Correlation immunity plays an important role in the context of correlation attacks on stream ciphers based on the combiner model~\cite{siegenthaler85}. More recently, this criterion also gained relevance for designing masking countermeasures to withstand side-channel analysis. In this case, the goal is to find a $t$-th order correlation immune function to resist SCA attacks of order $t$. At the same time, it is desirable that this function has the lowest Hamming weight possible, to have an efficient implementation of the masking countermeasure.

Beside the Walsh transform, correlation immune functions have also a nice combinatorial characterization in terms of orthogonal arrays. Formally, an orthogonal array of $N$ runs, $k$ factors, $s$ levels and strength $t$ (denoted as an $OA(N,k,s,t)$) is a $N\times n$ array with entries from a set $S$ with $s$ elements such that, for any $N\times t$ subarray, each $t$-uple of $S^t$ occurs exactly $\lambda = N/s^t$ times~\cite{hedayat99}. The value $\lambda$ is also called the index of the OA, and is completely determined by the other parameters. The link between binary OA (i.e. with $s=2$ levels) and correlation immune functions is given by the following result proved in~\cite{camion91}:
\begin{lemma}
\label{lm:oa-ci}
A Boolean function $f: \F_2^n \to \F_2$ is correlation immune of order $t$ if and only if its support $supp(f) = \{x \in \F_2^n: f(x) \neq 0\}$ is an $OA(N,n,2,t)$.
\end{lemma}
In other words, one can reduce the design of $n$-variable, $t$-th order correlation immune Boolean functions for SCA masking countermeasures to the search of binary OA of $n$ factors and strength $t$. The requirement of minimizing the Hamming weight of the function corresponds to the minimization of the number of runs $N$ in the OA. Once such an OA has been found, one can define the corresponding correlation immune function $f$ by taking the runs of the OA as the vectors in the support of $f$.

\subsection{Cellular Automata and Latin Squares}
\label{subsec:ca-ls}
In this work, we use cellular automata (CA) as a particular kind of vectorial Boolean functions, which we formally define below:
\begin{definition}
\label{def:ca}
Let $n,d \in \N$ with $d \le n$, and let $f: \F_2^d \to \F_2$ be a $d$-variable Boolean function. A cellular automaton of length $n$, diameter $d$ and local rule $f$ is a mapping $F: \F_2^n \to \F_2^{n-d+1}$ defined for all $x \in \F_2^n$ as:
\begin{equation}
\label{eq:ca}
F(x_1,\cdots,x_n) = (f(x_1,\cdots, x_d),f(x_2,\cdots, x_{d+1}),\cdots,f(x_{n-d+1},\cdots,x_n)) \enspace .
\end{equation}
\end{definition}
Intuitively, each output coordinate $i \in \{1,\cdots,n-d+1\}$ of the CA $F$ is determined by evaluating $f$ on the local neighborhood $(x_i,\cdots,x_{i+d-1})$. Remark that the output is smaller than the input: indeed, we can apply $f$ as long as there are enough neighboring coordinates to the right of the current output cell $i$. Therefore, we do not enforce any boundary conditions on the CA state, as it is commonly done in the CA literature. This means that the CA cannot be iterated
for multiple time steps, but this issue does not concern us since we are interested only in the single-step application of $F$. This model is also called no-boundary CA in related works~\cite{mariot19,mariot20}.

We now introduce the basic concepts related to Latin squares to recall the main results of~\cite{mariot20}. For all $n \in \N$, let us denote by $[n] = \{1,\cdots, n\}$ the set of the first $n$ natural numbers. A Latin square of order $n$ is a $n \times n$ matrix $L$ such that each row and each column of $L$ is a permutation of $[n]$. Equivalently, this means that each number from $1$ to $n$ occurs exactly once in each row and in each column of $L$. Then, two Latin squares $L_1,L_2$ of order $n$ are called orthogonal if their superposition yields all pairs in the Cartesian product $[n] \times [n]$. Formally, for any distinct pairs of coordinates $(i_1,j_1), (i_2,j_2) \in [n] \times [n]$, one has:
\begin{equation}
\label{eq:ols}
(L_1(i_1,j_1),L_2(i_1,j_1)) \neq (L_1(i_2,j_2),L_2(i_2,j_2)) \enspace .
\end{equation}
A set of $k$ Latin squares $L_1,\cdots,L_k$ of order $n$ that are pairwise orthogonal is also called a set of $k$-MOLS (mutually orthogonal Latin squares). The construction of MOLS is a rich research line in the combinatorial designs literature, and finds several applications in cryptography, coding theory and statistics~\cite{stinson04}. Interestingly, $k$-MOLS of order $n$ are also equivalent to orthogonal arrays with $N = n^2$ runs, $k$ factors, $n$ levels and strength $2$. Indeed, one can construct an $OA(N,k,n,2)$ from a set of $k$-MOLS by ``linearizing'' each Latin square as a column of the OA: for each $i \in [k]$, the $i$-th column of the OA corresponds to the Latin square $L_i$ in the MOLS set, with its entries listed in lexicographic order.

The authors of~\cite{mariot20} proposed a method to construct sets of $k$-MOLS from cellular automata, by focusing on the subclass of bipermutive rules. Formally, a local rule $f: \F_2^d \to \F_2$ is called bipermutive if it is defined as $f(x_1,\cdots, x_d) = x_1 \oplus \varphi(x_2,\cdots,x_{d-1}) \oplus x_d$ for all $x \in \F_2^d$, where $\varphi: \F_2^{d-2} \to \F_2$ is any function of the $d-2$ central variables. Then, one can construct a Latin square $L_F$ of order $n = 2^{d-1}$ from a CA $F: \F_2^{2(d-1)} \to \F_2^{d-1}$ as follows:
\begin{compactenum}
\item The left half of the CA input $(x_1,\cdots,x_{d-1})$ is used to index the rows of $L_F$.
\item The right half $(x_d,\cdots,x_{2(d-1)})$ is used to index the columns of $L_F$.
\item The output $F(x_1,\cdots,x_{2(d-1)})$ of the CA is used as the entry indexed respectively by the coordinates $(x_1,\cdots,x_{d-1})$ and $(x_d,\cdots,x_{2(d-1)})$.
\end{compactenum}
Clearly, the procedure above assumes that a bijective mapping $\phi: \F_2^{d-1} \to [n]$ is used to convert the binary vectors of $\F_2^{d-1}$ in numbers from $1$ to $2^{d-1}$, and vice versa. The authors of~\cite{mariot20} focused on the construction of MOLS from bipermutive CA by further focusing on the class of linear rules. Here, however, we will consider the general setting of MOLS defined by generic bipermutive CA. In particular, we define two bipermutive CA $F_1,F_2: \F_2^{2(d-1)} \to \F_2^{d-1}$ to be orthogonal (or equivalently they are OCA) if the corresponding Latin squares are orthogonal. Accordingly, a family of $k$ pairwise orthogonal bipermutive $CA$ $F_1,\cdots, F_k$ will be called a set of $k$-MOCA (mutually orthogonal cellular automata).

\section{Construction of Correlation Immune Functions}
\label{sec:constr}
We now prove our main result, namely that a set of $k$-MOCA can be used to define a binary OA of strength at least $2$. This will allow us, in turn, to construct correlation immune functions of order at least $2$. To this end, let us first review the concept of coupled de Bruijn graph introduced in~\cite{mariot18}, which will be useful in our proof.

Recall that a de Bruijn graph of order $b$ over a set $S$ of $m$ symbols is defined as $G_{m,b} = (V,E)$, where $V = S^b$, while two vertices $u,v \in V$ are connected by a directed edge if and only if they overlap respectively on the rightmost and leftmost $b-1$ coordinates. Assume now that $S = \F_2$, and let $b = d-1$. A local rule $f: \F_2^d \to \F_2$ of diameter $d$ can be represented as a labeling function $l_f: E \to \F_2$ on the edges of $G_{2,b}$. In particular, for each pair $u,v \in E$, we set $l(u,v) = f(u \odot v)$, where $u \odot v \in \F_2^d$ represents the fusion of $u$ and $v$ as defined in~\cite{sutner91}. In other words, $u \odot v$ is the $d$-variable vector formed by adding the last coordinate of $v$ to $u$. Then, it can be seen that the output of a CA equipped with rule $f$ corresponds to a path on the edges of the associated de Bruijn graph, following the overlapping vertices that form a particular input. Remark that if $f$ is bipermutive, then the labels of the outgoing (respectively, ingoing) edges of any vertex $v \in V$ form a permutation of $\F_2$. This implies that one can traverse the edges in both directions, and that the CA is surjective.

Suppose now that we have two labelings $l_f,l_g: E \to \F_2$ defined by bipermutive rules $f,g: \F_2^d \to \F_2$. We call the corresponding graph as the coupled de Bruijn graph associated to $f$ and $g$, since each label is now a pair of bits $(f(u \odot v), g(u \odot v))$ for each edge $(u,v) \in E$. We call two bipermutive labelings $l_f,l_g$ orthogonal if for each pair of vectors $(x,y) \in \F_2^{b}$ there exists exactly one path on the edges of the coupled de Bruijn graph that is labelled by $(x,y)$. It is not difficult to see that this is an equivalent characterization of orthogonal CA. We formally state this fact below, since it will become useful later in our proof:
\begin{lemma}
\label{lm:ort-lab}
Let $d \in \N$ with $b = d-1$, and $f,g: \F_2^d \to \F_2$ be two bipermutive rules of diameter $d$. Then, the CA $F,G: \F_2^{2b} \to \F_2^b$ respectively equipped with rule $f$ and $g$ are orthogonal if and only if the labelings on the coupled de Bruijn graph of $f$ and $g$ are orthogonal.
\end{lemma}

To fix ideas, let $d=3$, and assume that the two bipermutive local rules are $f(x_1,x_2,x_3) = x_1 \oplus x_3$ and $g(x_1,x_2,x_3) = x_1 \oplus x_2 \oplus x_3$ (respectively rules 90 and 150 in Wolfram's notation). The two rules induce orthogonal CA with Latin squares of order $2^2 = 4$, and the corresponding coupled de Bruijn graph is depicted in Figure~\ref{fig:example}.
\begin{figure}[t]
\begin{minipage}{0.5\textwidth}
\centering
\resizebox{!}{5cm}{%
\Large
\begin{tikzpicture}
[->,auto,node distance=1.5cm, every loop/.style={min distance=12mm},
       empt node/.style={font=\sffamily,inner sep=0pt,outer sep=0pt},
       circ node/.style={circle,thick,draw,font=\sffamily\bfseries,minimum
         width=0.8cm, inner sep=0pt, outer sep=0pt}]

       \node [empt node] (e1) {};
       \node [circ node] (n00) [above=1.75cm of e1] {$00$};
       \node [circ node] (n01) [right=1.75cm of e1] {$01$};
       \node [circ node] (n10) [left=1.75cm of e1] {$10$};
       \node [circ node] (n11) [below=1.75cm of e1] {$11$};       

       \draw [->, thick, shorten >=0pt,shorten <=0pt,>=stealth] (n00) 
       edge[bend left=20] node (f5) [above right]{$1,1$} (n01);
       \draw [->, thick, shorten >=0pt,shorten <=0pt,>=stealth] (n01)
       edge[bend left=20] node (f5) [below right]{$1,0$} (n11);
       \draw [->, thick, shorten >=0pt,shorten <=0pt,>=stealth] (n11)
       edge[bend left=20] node (f5) [below left]{$1,0$} (n10);
       \draw [->, thick, shorten >=0pt,shorten <=0pt,>=stealth] (n10)
       edge[bend left=20] node (f5) [above left]{$1,1$} (n00);
       \draw[->, thick, shorten >=0pt,shorten <=0pt,>=stealth] (n10) edge[bend
         left=20] node (f1) [above]{$0,0$} (n01);
       \draw[->, thick, shorten >=0pt,shorten <=0pt,>=stealth] (n01)
       edge[bend left=20] node (f2) [below]{$0,1$} (n10);
       \draw[->, thick, shorten >=0pt,shorten <=0pt,>=stealth] (n00) edge[loop
         above] node (f3) [above]{$0,0$} ();
       \draw[->, thick, shorten >=0pt,shorten <=0pt,>=stealth] (n11) edge[loop
         below] node (f4) [below]{$0,1$} ();
\end{tikzpicture}
}
\end{minipage}%
\begin{minipage}{0.5\textwidth}
\centering
    \begin{tabular}{c|cc}
                \hline
		$(v_1,v_2) \to (u_1,u_2)$ & $l_f$     & $l_g$ \\ 
                \hline
	  $00 \to 00$            & 0            &   0      \\
      $10 \to 00$            & 1            &   1      \\
      $01 \to 10$            & 0            &   1      \\
      $11 \to 10$            & 1            &   0      \\
      $00 \to 01$            & 1            &   1      \\
      $10 \to 01$            & 0            &   0      \\
      $01 \to 11$            & 1            &   0      \\
      $11 \to 11$            & 0            &   1      \\
      
\hline
    \end{tabular}
    \end{minipage}
\caption{Example of orthogonal labelings for the de Bruijn graph $G_{2,2}$ induced by the CA local rules 90 and 150 of diameter $d=3$.}
\label{fig:example}
\end{figure}
One can see that the two labelings $l_f,l_g$ defined in the table are indeed bipermutive and orthogonal. In particular, for each of the $16$ pairs $(x,y) \in (\F_2^2)^2$ there is exactly one path of length $2$ in $G_{2,2}$ labeled by $(x,y)$.

Let $F_1, F_2, \cdots, F_k: \F_2^{2b} \to \F_2^{b}$ be a set of $k$-MOCA respectively defined by local rules $f_1,\cdots, f_k: \F_2^d \to \F_2$ of diameter $d$, with $b=d-1$. We define the $N \times n$ array $A$ where $N = 2^{2b}$ and $n = kb$ as follows: for each $(x,y) \in \F_2^{2b}$, the row of $A$ indexed by $(x,y)$ is equal to:
\begin{equation}
\label{eq:moca-oa}
A(x,y) = (F_1(x,y), F_2(x,y), \cdots, F_k(x,y)) \enspace .
\end{equation}
In other words, $A$ is formed by simply juxtaposing the output of the $k$ MOCA for each possible combination of input $(x,y) \in \F_2^{2b}$. We now prove that this array is an OA of strength $2$.
\begin{lemma}
\label{lm:moca-oa}
The array $A$ defined in Eq.~\eqref{eq:moca-oa} is an $OA(N, n, 2, 2)$, where the number of runs is $N = 2^{2b}$ and the number of factors is $n=kb$.
\begin{proof}
We need to show that in any subset of $t=2$ columns $i,j$ of $A$ each pair of bits $(x_i,x_j) \in \F_2^2$ occurs exactly $\lambda = N/2^t = 2^{2b-2}$ times. Without loss of generality, we can assume that $k=2$, since $F_1, \cdots F_k$ is a family of $k$-MOCA. Hence, we have two main cases to check for two columns $i\neq j$:

\begin{compactenum}
\item $i,j$ belong to the output of the same CA $F_l$.
\item $i,j$ belong to the output of two different CA $F_l,F_m$ (which are orthogonal).
\end{compactenum}

Let us start from the first case, i.e. $i$ and $j$ are chosen among the columns of the same CA $F_l$. Let $(\tilde{x}_i, \tilde{x}_j) \in \F_2^2$ be the value of the two bits of which we want to compute the multiplicity of occurrence in columns $i$ and $j$. Since the two CA $F_l$ and $F_m$ are orthogonal, it means that by fixing the output of $F_m$ to a specific vector $(y_1,\cdots,y_b) \in \F_2^b$, each possible vector $(x_1,\cdots, x_b) \in \F_2^b$ occurs exactly once as an output of $F_l$. Suppose now that we fix the $i$-th and $j$-th coordinates respectively to $\tilde{x}_i$ and $\tilde{x}_j$ in the output of $F_l$. Since we have $2^{b-2}$ free coordinates, it follows that the pair $(\tilde{x}_i, \tilde{x}_j)$ occurs $2^{b-2}$ times if we keep the output of $F_m$ fixed to $(y_1,\cdots,y_b)$. If we consider the occurrences of $(\tilde{x}_i, \tilde{x}_j)$ in $F_l$ across all possible outputs of $F_m$ we need to multiply $2^{b-2}$ by $2^b$, i.e. the number of possible ways to fix the output of $F_m$. Therefore the total number of occurrences is $2^{b-2} \cdot 2^b = 2^{2b-2}$.

Suppose now that $i$ and $j$ are columns respectively of $F_l$ and $F_m$. If all output coordinates of $F_l$ and $F_m$ are fixed respectively to $x$ and $y \in \F_2^b$, then there exists a single row of $A$ labeled by $x$ and $y$, since $F_l$ and $F_m$ are orthogonal, and by Lemma~\ref{lm:ort-lab} there is a unique path on the coupled de Bruijn graph labelled by $(x,y)$. We proceed by induction on the number of free coordinates in $(x,y)$ to show that if we only have two of them fixed, i.e. $\tilde{x}_i$ and $\tilde{y}_j$, then there are exactly $2^{2b-2}$ paths on the de Bruijn graph that feature $\tilde{x}_i$ and $\tilde{y}_j$ in those coordinates. As a base case, suppose that we have only one free coordinate in the pair of paths, i.e. all other $2b-1$ are fixed. Then, since each of the two labelings is bipermutive, it follows that there are exactly 2 paths labelled by the $2b-1$ fixed coordinates. For the induction step, suppose that there are $ 1 \le p < 2b$ free coordinates, and thus $2^{p}$ paths labelled by the remaining $2b-p$ fixed coordinates by induction hypothesis. If we free an additional coordinate, we need to multiply the number of paths with $p$ free coordinates by 2, since each of them can be completed in 2 different ways in the additional free coordinate, due to the bipermutivity of the two rules. Hence, the number of partially labelled paths with $p+1$ free coordinates is $2^{p+1}$. If we take the particular case where only 2 coordinates $i$ and $j$ are fixed respectively to $\tilde{x}_i$ and $\tilde{y}_j$ (or equivalently, $2b-2$ are free), it follows that there are $2^{2b-2}$ paths partially labeled by $\tilde{x}_i$ and $\tilde{y_j}$.\qed
\end{proof}
\end{lemma}

Putting together Lemma~\eqref{lm:oa-ci} and~\ref{lm:moca-oa}, we have thus obtained the following result to construct a second-order correlation immune function from a set of $k$-MOCA:
\begin{theorem}
\label{thm:ci-moca}
Let $F_1,\cdots, F_k: \F_2^{2b} \to \F_2^b$ be a set of $k$-MOCA of diameter $d = b+1$, and let $n = kb$. Then, the $n$-variable function $f: \F_2^n \to \F_2$ whose support is defined by the array $A$ in Eq.~\eqref{eq:moca-oa} is correlation immune of order at least $2$. In particular, the Hamming weight of $f$ is $N=2^{2b}$.
\end{theorem}

Remark that the case where $k=2$ trivially gives the constant function $f(x) = 1$ for all $x \in \F_2^n$. As a matter of fact, the number of input variables is $n=kb = 2b$, which means that the truth table of $f$ is composed of $2^{2b}$ values. At the same time, the number of runs of the OA corresponding to $k=2$ MOCA is also $N=2^{2b}$. Therefore, the support of $f$ coincides with its whole truth table. For this reason, in what follows we will address mainly the case where $k=3$.

\section{Computational Search Results}
\label{sec:search}
To investigate the construction described in the previous section, we performed an exhaustive search of all $k$-MOCA for $k=3$ and $d=4,5$. In particular, we discarded $d=3$ since there are no families of $3$-MOCA in that case. On the other hand, going for higher values of $k$ and $d$ makes the search space too huge to be exhaustively visited in a limited time. Following Theorem~\ref{thm:ci-moca}, this means that we addressed the construction of correlation immune functions of $n=9,12$ variables.

We first generated all pairs of OCA (i.e., $2$-MOCA) of diameters $d=4,5$ by using the combinatorial algorithm described in~\cite{mariot17}. Then, we incrementally constructed the families of $3$-MOCA by exhaustively visiting each of the $2^{2^{d-2}}$ bipermutive rules of diameter $d$, and checking that the corresponding Latin squares were orthogonal with each of those in the previously generated lists of OCA. Next, we defined the corresponding Boolean functions of $n=3b$ variables using Theorem~\ref{thm:ci-moca}, and computed their Walsh transform to verify their order of correlation immunity. Table~\ref{tab:res} reports the main results of this computational search experiment. In particular, for each considered diameter ($d$) the table gives the number of $3$-MOCA generated (\#$3$-MOCA), the number of variables ($n$), the Hamming weight ($w_H$), the correlation immunity order ($CI$) and the number of functions achieving that order (\#$CI$), and the best known lower bound on the Hamming weight for the corresponding order of correlation immunity (Min($w_H$)), taken from~\cite{carlet21}.

\setlength{\tabcolsep}{0.8em}
\begin{table}[t]
	\centering
	\caption{Classification of correlation immune functions generated by $3$-MOCA of diameter $d \in \{4,5,6\}$.}
	\begin{tabular}{ccccccc}
		\hline\noalign{\smallskip}
		$d$ & \#$3$-MOCA & $n$ & $w_H$ & $CI$ & \#$CI$ & Min$w_H$\\
		\noalign{\smallskip}\hline\noalign{\smallskip}
		\hline\noalign{\smallskip}
		4 & 2 & 9 & 64 & 3 & 2 & 20 \\
		\hline\noalign{\smallskip}
		5 & 36 & 12 & 256 & 3 & 27 & 24 \\
		5 & 36 & 12 & 256 & 4 & 6 & 24 \\
		\hline\noalign{\smallskip}
	\end{tabular}
	\label{tab:res}
\end{table}

From the table, one can see that all generated functions actually have a correlation immunity order higher than $2$. Indeed, both functions of $n=9$ variables generated from the $3$-MOCA of diameter $d=4$ have correlation immunity order $3$, as the majority of functions of $n=12$ functions obtained from $3$-MOCA of diameter $d=5$. Further, a few functions of $n=12$ variables obtained from this latter case are even $4$-th order correlation immune. This empirical observation suggests that the correlation immunity order proved in Theorem~\ref{thm:ci-moca} may not be a tight lower bound. Further experiments on larger diameters should be performed to see if this hypothesis holds, or if there exist functions constructed through our methods that are effectively second-order correlation immune. Additionally, one can remark that the Hamming weight of the functions generated through our method are far from the known best lower bounds, reported in the last column of Table~\ref{tab:res}. Indeed, for $n=9$ variables we obtain functions of weight $64$, while the best lower bound is 20. For $n=12$ variables the gap is even greater, since the weight of our functions is $256$, while the lower bound is $24$.

\section{Conclusions}
\label{sec:outro}
In this work, we proposed a method to construct correlation immune Boolean functions from sets of mutually orthogonal cellular automata. Our main theoretical result shows that the binary array formed by juxtaposing the output tables of a set of $k$ MOCA is an orthogonal array of strength $2$. Hence, on account of Lemma~\ref{lm:oa-ci} such an array can be used as the support of a Boolean function with correlation immunity order at least $2$. Our exhaustive search experiments on the sets of $3$-MOCA defined by rules of diameters $d=4,5$ show an interesting fact, namely that the resulting Boolean functions always have a higher order of correlation immunity, namely at least $3$.

There are several directions along which this work can be extended in future research. The most natural open question remains whether our Theorem~\ref{thm:ci-moca} is really a tight lower bound on the correlation immunity of the Boolean functions constructed through $k$-MOCA. If further experiments show that the lowest correlation immunity order is always at least $3$, then it would be interesting to try refining Lemma~\ref{lm:moca-oa}, and prove that the binary OA obtained from a set of $k$-MOCA has always at least strength $3$.

A second direction concerns the non-optimal weights of the correlation immune functions generated by our method. Indeed, one possible approach to improve on this aspect would be to adopt the so-called expurgation procedure used in the field of error-correcting codes. In the context of orthogonal arrays, this basically amounts to the removal of a subset of rows, such that the resulting array is still an OA of the same strength (but clearly with a smaller index $\lambda$). The choice of the rows to be removed can be conceived as a combinatorial optimization problem, which could be addressed with different optimization algorithms.

\bibliographystyle{abbrv}
\bibliography{bibliography}

\end{document}